# Growth of GeO$_2$ on *R*-plane and *C*-plane Sapphires by MOCVD

Imteaz Rahaman[1], Hunter D. Ellis[1], Kathy Anderson[2], Michael A. Scarpulla[1], Kai Fu[1, a)]

Department of Electrical and Computer Engineering, The University of Utah, Salt Lake City, UT 84112, USA

Utah Nanofab, Price College of Engineering, The University of Utah, Salt Lake City, UT 84112, USA

a) E-mail: kai.fu@utah.edu

**Abstract:** Rutile Germanium Dioxide (GeO$_2$) has been recently theoretically identified as an ultrawide bandgap (UWBG) semiconductor with bandgap 4.68 eV similar to Ga$_2$O$_3$ but having bipolar dopability and ~2x higher electron mobility, Baliga figure of merit (BFOM) and thermal conductivity than Ga$_2$O$_3$. Bulk crystal growth is rapidly moving towards making large sized native substrates available. These outstanding material properties position GeO$_2$ as a highly attractive UWBG semiconductor for various applications. However, the epitaxial growth in the most advantageous polymorph (rutile), ensuring controlled phase, pristine surface/interface quality, precise microstructure, and optimal functional properties, is still in its infancy. In this work, we explored growth of GeO$_2$ by metal-organic chemical vapor deposition (MOCVD) on both *C*- and *R*-plane sapphire. Utilizing tetramethylgermane (TMGe) as a precursor, we have investigated the influences of different parameters on the film properties, including growth temperature, chamber pressure, TMGe flow rate, oxygen flow rate, shroud gas flow rate, and rotation speed. The total pressure emerged as a crucial parameter while growth attempts at low total pressure resulted in no films for a wide range of temperatures, precursor flow rate, argon flow rates, and susceptor rotation rate. A phase diagram, derived from our experimental findings, delineates the growth windows for GeO$_2$ films on sapphire substrates. This study serves as a pioneering guide for the MOCVD growth of GeO$_2$ films.

1. Introduction:

Ultrawideband gap (UWBG) semiconductors such as diamond, AlN, and Ga$_2$O$_3$ are becoming increasingly important in scientific and technological research due to their excellent electrical, optical, thermal, and chemical stability [1], [2]. These characteristics make them highly promising candidates for fulfilling the evolving needs of current and future technologies, particularly in the realm of minimizing power loss and subsequently reducing energy-related emissions [1], [3], [4]. However, the utilization of these UWBG materials is hindered by challenges associated with doping and/or the production of large-sized wafers. Germanium dioxide (GeO$_2$), a newly identified ultrawide bandgap (UWBG) semiconductor, exhibits a notable 4.68 eV bandgap, supports both n-type and p-type doping, and surpasses Ga$_2$O$_3$ more than 2 times in both Baliga Figure of Merit (BFOM) and thermal conductivity [5], [6], [7]. These advanced characteristics, along with the availability of large single-crystal substrates, make GeO$_2$ highly promising for high-power electronics, deep-UV optoelectronics/photonics, and energy storage applications [5], [8], [9], [10]. $GeO_2$ primarily exists in two phases: α-quartz with a hexagonal structure, and rutile with a tetragonal structure. Both forms of $GeO_2$ are categorized as ultrawide bandgap (UWBG) semiconductors, featuring a bandgap well over 4 eV [11], [12], [13]. Due to its higher carrier mobility (> 244 $cm^2V^{-1}s^{-1}$) and superior ambipolar dopability, $GeO_2$ emerges as an ideal material for creating efficient, next-generation electronic and optoelectronic devices through homoepitaxial growth of p-n junctions [7], [14]. Additionally, $GeO_2$ boasts higher intrinsic thermal conductivity (37 $Wm^{-1}K^{-1}$ along the *a*-direction and 58 $Wm^{-1}K^{-1}$ along the *c*-direction at 300 K) compared to $Ga_2O_3$, underscoring its potential in high-performance applications [15]. However, despite its promising attributes for a wide range of applications, synthesizing $GeO_2$ thin films having the desired crystal phase is still in its infancy, presenting a notable obstacle. This challenge is primarily due to the polymorphism of $GeO_2$, where the similar free energies of its polymorphs (including amorphous) lead to a propensity for deep metastable states. For any growth method, this complicates the discovery of the stable thermodynamic and kinetic window for growing the desired r-$GeO_2$. The formation of the glass phase of $GeO_2$ is the most straightforward near room temperature [14], [16]. This phase is the one famously soluble in water. The hexagonal (quartz) $GeO_2$ unit cell is ~40% smaller in volume than its tetragonal rutile counterpart, further complicating the phase stabilization of material during processing [17]. As a result, complex or high-pressure techniques (over 100 MPa) are often necessary to synthesize certain $GeO_2$ polymorphs [17], [18].

Recently, several research groups have made advancements in $GeO_2$ growth. In 2020, Chae *et al.* successfully epitaxially stabilized r-$GeO_2$ thin films on *R*-plane sapphire substrates with a buffer layer of $(Sn,Ge)O_2$ using molecular beam epitaxy (MBE) [14]. The work also shows a narrow growth window for r-$GeO_2$. In 2021, Zhou *et al.* delve into solid-state crystallization, highlighting the nuanced interplay between temperature, nucleation, and substrate effects on the growth morphology of $GeO_2$, transitioning from spherulitic to single crystalline structures with increased annealing temperature [19]. Meanwhile, Deng *et al.* reported pulsed laser deposition to synthesize $GeO_2$ films, facing the daunting challenge of stabilizing the rutile phase amidst metastable glass forms [8]. Takane *et al.* established a growth route via mist chemical vapor deposition (mist CVD) in 2021, achieving growth rates of 1.2–1.7 µm/h on (001) r-$TiO_2$ substrates [20]. In 2022, Gaurav *et al.* explored radio-frequency magnetron sputtering methods, showcasing the ability to fine-tune optical properties amidst the challenges of achieving phase purity due to the closely matched free energy of polymorphs of $GeO_2$ [9]. This difficulty manifests in the coexistence of nanotextured hexagonal and tetragonal phases. Despite the exciting progress in growing thin film rutile $GeO_2$, there have been no reports of MOCVD based growth, while MOCVD is the prevailing method so far for large-scale wafer epitaxy in current mass production processes and typically offers growth rates of order a hundred times faster than MBE using elemental sources (but on par with suboxide sources) [21].

In this study, we have investigated growing $GeO_2$ films on *C*- and *R*-plane sapphires by MOCVD. Various parameters, including growth temperature, chamber pressure, shroud gas, oxygen flow rate, and susceptor rotation speed, have been explored to identify the optimal conditions for crystalline $GeO_2$ formation. Films were characterized by X-ray Diffraction (XRD), Scanning

Electron Microscopy (SEM), Energy Dispersive X-ray Spectroscopy (EDX), and Atomic Force Microscopy (AFM). We map out a phase diagram delineating the transition of GeO$_2$ from amorphous to polycrystalline for MOCVD growth.

## 2. Experimental Section

The growth was conducted in the Agilis MOCVD system by Agnitron Technology. The GeO$_2$ films with thickness ranging from 0.73 µm to 2.1 µm were grown for 90 minutes (min) on both *R*-plane and *C*-plane sapphire substrates for comparison by varying the growth temperature from 725°C to 925°C and the chamber pressure from 100 Torr to 300 Torr. The thickness of the film is measured by a surface profilometer from the distinguished region of substrate with no film to the spherulites or amorphous region. Tetramethylgermane (TMGe) and pure oxygen (O$_2$) were used as the precursors while argon (Ar) serves as the carrier gas and shroud gas. The flow rate of O$_2$ varied from 450 to 1800 SCCM, the shroud gas (Ar) flow rate from 400 SCCM to 2800 SCCM, and the TMGe precursor flow rate from 3.5 SCCM to 10 SCCM. The susceptor rotation speed was either 300 RPM or 2 RPM to explore the impact of the boundary layer thickness. The bubbler temperature used for each growth was 3 °C, with a specified cooling rate of approximately 30 °C/min. Before loading the sapphire substrates into the MOCVD chamber, the sapphire substrates were cleaned using piranha solution (H$_2$SO$_4$:H$_2$O$_2$ = 3:1), followed by acetone, isopropanol, and DI water. The list of growth parameters, along with their corresponding sample IDs, has been summarized in Table I.

**Table I.** *Summary of GeO$_2$ films grown at different growth conditions, including growth temperature, chamber pressure, TMGe precursor flow rate, O$_2$ flow rate, shroud gate flow rate, and susceptor rotation speed.*

| Sample ID | Temperature (°C) | Pressure (Torr) | Precursor Flow rate (SCCM) | Oxygen Flow rate (SCCM) | Shroud gas Flow rate (SCCM) | Rotation speed (RPM) | Thickness (µm) |
|---|---|---|---|---|---|---|---|
| E1  | 925 | 300 | 10  | 1800 | 2800 | 300 | 2.1 |
| E3  | 825 | 300 | 3.5 | 1800 | 2800 | 300 | 0.91 |
| E4  | 725 | 300 | 3.5 | 1800 | 2800 | 300 | 0.73 |
| E5  | 925 | 200 | 3.5 | 1800 | 2800 | 300 | 1.07 |
| E6  | 925 | 100 | 3.5 | 1800 | 2800 | 300 | No film |
| E7  | 925 | 300 | 3.5 | 1800 | 2800 | 300 | 1.77 |
| E9  | 925 | 300 | 3.5 | 900  | 2800 | 300 | 1.68 |
| E10 | 925 | 300 | 3.5 | 450  | 2800 | 300 | 1.59 |
| E12 | 925 | 300 | 3.5 | 900  | 1400 | 300 | 1.62 |
| E15 | 925 | 300 | 3.5 | 900  | 400  | 300 | 1.53 |
| E16 | 925 | 300 | 10  | 1800 | 2800 | 2   | 0.76 |

A Philips PANalytical X-Pert X-ray Diffractometer (XRD) was used to determine the structural properties of the GeO$_2$ thin film. Micron-scale EDS Quanta 600F environmental SEM was used to analyze the surface morphology and chemical composition of the surface of the films. The Bruker Dimension ICON AFM was used to study the surface morphology and roughness of the film surface.

## 3. Results and Discussions

In this section, we will explore how variations in growth temperature, chamber pressure, oxygen flow rate, shroud gas flow rate, and rotation speed influence the structure and surface morphology of the film, providing a comprehensive analysis of their effects.

### 3.1 Effects of growth temperature

The effects of varying growth temperatures on the crystal structure of $GeO_2$ thin films are delineated in Figures 1 and 2. Figures 1(a) and (b) showcase the XRD patterns of $GeO_2$ thin films grown on *R*-plane and *C*-plane sapphire substrates, respectively, with different growth temperatures and other growth parameters detailed in Table I for sample IDs E7, E3, and E4. The XRD reveals a transition of the $GeO_2$ films from an amorphous phase at 725 °C to a polycrystalline quartz phase at 925 °C. In the analysis of the polycrystalline quartz phase, distinct X-ray diffraction peaks were identified from films deposited on *R*-plane sapphire substrates at the (100), (-1-10), (110), (012), (220), (112), (121), (113), (212), (301), and (032) planes. Similarly, all these peaks are also seen on the *C*-plane sapphire substrate along with (101), (112), (202), (301), and (024) planes. The additional peaks specific to the *C*-plane are not observed on the *R*-plane due to the overlapping with *R*-plane sapphire peaks. Notably, the (220) plane is exclusive to the *R*-plane, absent on the *C*-plane due to similar overlapping issues. All the peaks agree with XRD patterns for the polycrystalline quartz $GeO_2$. Additionally, it is important to note that the peaks of rutile $GeO_2$ may overlap with those of quartz $GeO_2$. At a temperature of 725 °C, no distinct peaks for quartz $GeO_2$ but an amorphous hump are observed, indicating that the film is amorphous. However, the film becomes polycrystalline when the temperature increases to 825 °C and 925 °C. This change can be attributed to the increased mobility of adatoms at higher temperatures, which is insufficient at lower temperatures, preventing the film from crystallizing [14]. Additionally, Takane *et al.* have emphasized the importance of balancing adsorption and desorption to successfully form a crystalline film, regardless of the growth mechanism, such as Molecular Beam Epitaxy (MBE) or mist Chemical Vapor Deposition (CVD) [20]. In fact, achieving a balance between adsorption and desorption helps distinguish between growth and etching conditions: Growth rate (GR) > 0 indicates net growth while GR < 0 indicates net etching. To differentiate between amorphous and crystalline structures, the comparison of net mass addition rate (growth rate) to the adatom diffusivity can display whether the deposited layer will be amorphous or crystalline. This balance may also be crucial for our MOCVD method to achieve crystallization following our specific recipe. Furthermore, the intensity of the quartz $GeO_2$ peak increases with higher growth temperatures for both the *C*-plane and *R*-plane sapphire substrate. As the film thickness grows— 0.73 µm at 725 °C, 0.91 µm at 825 °C, and 1.77 µm at 925 °C—the intensity of the peak also rises in line with this increased thickness. Figure 1(c) shows that the growth rate of the film increases with rising temperatures: 0.49 µm/h at 725 °C, 0.61 µm/h at 825 °C, and 1.18 µm/h at 925 °C. This increase in the growth rate can be explained by two possible factors. First, the diffusion of precursor gases and the chemical reactions in the chamber or on the surface both accelerate as the temperature increases, which boosts the growth rate. Alternatively, if the chemical reaction rates reach their maximum and no longer increase, then the faster diffusion of precursor gases at higher temperatures becomes the main factor controlling the growth rate.

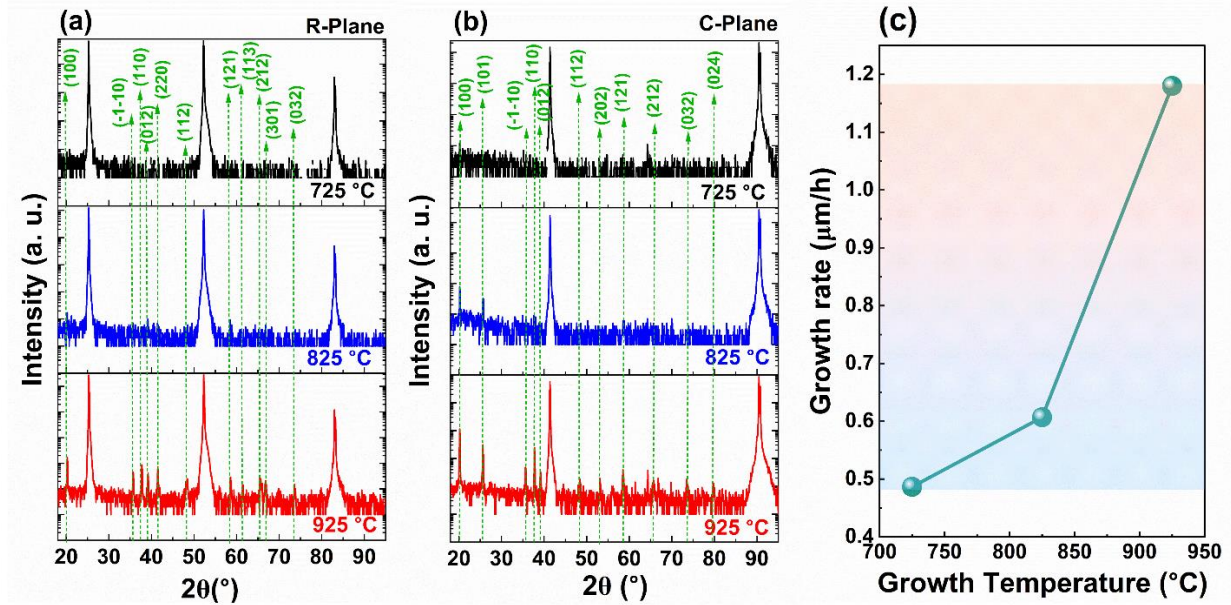

**Figure 1.** Impacts of different growth temperatures, ranging from 725 °C to 925 °C by XRD on (a) *R*-plane sapphire, (b) on *C*-plane sapphire substrate. The most prominent peaks in figures (a) and (b) represent the characteristic diffraction peaks from the *R*-plane and *C*-plane sapphire substrates, respectively. (c) Growth rate as a function of growth temperature.

The SEM images of Figure 2(a-c) and 2(g-i) clearly delineate the differences in surface morphology between the polycrystalline quartz and amorphous films on *R*-plane and *C*-plane sapphire substrates respectively. At a lower temperature of 725°C, the film remains amorphous, lacking any specific surface pattern. As we increase the temperature to 825°C, we begin to see a partial crystallization, though much of the material retains its amorphous state. Upon reaching 925°C, the crystalline quality markedly improves, covering nearly the entire surface. Spherulitic crystallization, which is prominent at 825 °C and 925 °C, arises due to a surge in lattice defects [22]. At the initial temperature rise to 825 °C, we observe the onset of dendritic growth, shifting to the formation of large quartz spherulites at 925 °C. These spherulites, typically 40-60 μm in diameter at 825 °C, expand to a significant 400-850 μm at the 925 °C temperature, as evident in Figures 2(a-b) and 2(g-h). Spherulites consist of fibers, spread out from a central core, growing in a non-regular pattern that often reveals misorientation angles within a 0-15° range [22]. A substantial crystallization driving force is essential to forming spherulites from a liquid phase. This driving force is crucial to overcoming the energy barriers associated with nucleation and the growth of organized crystal structures from the molten state [19]. At the core of each spherulite lies a flat region with a cracked center, evolving into a hexagonal shape that mirrors the natural symmetry of quartz. The spherulites preferentially nucleate in the center of the grains and appear lower in height than their amorphous surroundings. The spherulites vary in size and are marked by distinct boundaries where neighboring fibers display orientation angles that exceed 3° [19]. This variability in fiber orientation contributes to the complex and heterogeneous structure of the spherulites. With each SEM image, we gain insights into the fibrous composition of the grains, observing how their crystal direction subtly shifts during growth, a phenomenon indicative of the amorphous-to-crystalline transformation observed in other studies [23], [24]. Savytskii *et al.* suggest that the lattice rotation during growth is a result of dislocation rearrangements due to the stress of volume changes during crystallization [23]. Thermal stress also seems to play a critical role in determining lattice orientation. Besides spherulites observed in $GeO_2$ films can be compared to the formation of ice on a car windshield, where condensation from the vapor phase leads to crystallization. In this context, the interface energy plays a more prominent role than the volume energy, since the growth process is fundamentally a two-dimensional phenomenon. In the classical nucleation theory for three-dimensional systems, the surface area scales as $r^2$ while the volume scales as $r^3$, where $r$ is the characteristic size of the nucleated structure. However, in a two-dimensional system, both interface and volume parameters scale as $r^2$. This proportional scaling means that interface energy dominates in determining the morphology of the thin film, leading to the characteristic spherulitic patterns observed in the $GeO_2$ film. It's clear that temperature plays a crucial role in catalyzing the transformation and growth of $GeO_2$, transitioning it from an amorphous to a polycrystalline state with increasing temperature. EDX analyses corroborate the presence of $GeO_2$ with variations in Ge/O ratio at different temperatures, likely reflective of the changing film thickness. It can be noted that the Ge/O ratio is significantly lower at 725 °C, where the growth rate is 0.49 µm/h. This rate is much lower compared to the growth rates observed at 825 °C and 925 °C. For a more detailed understanding of the chemical composition, SEM-EDX analyses were conducted on two distinct areas on the *R*-plane sapphire at 825 °C: the spherulite regions and the smooth regions. Interestingly, the chemical compositions of both areas were found to be nearly identical and the shape of the spherulites is spherical, as clearly demonstrated in Figure 2(m-n).

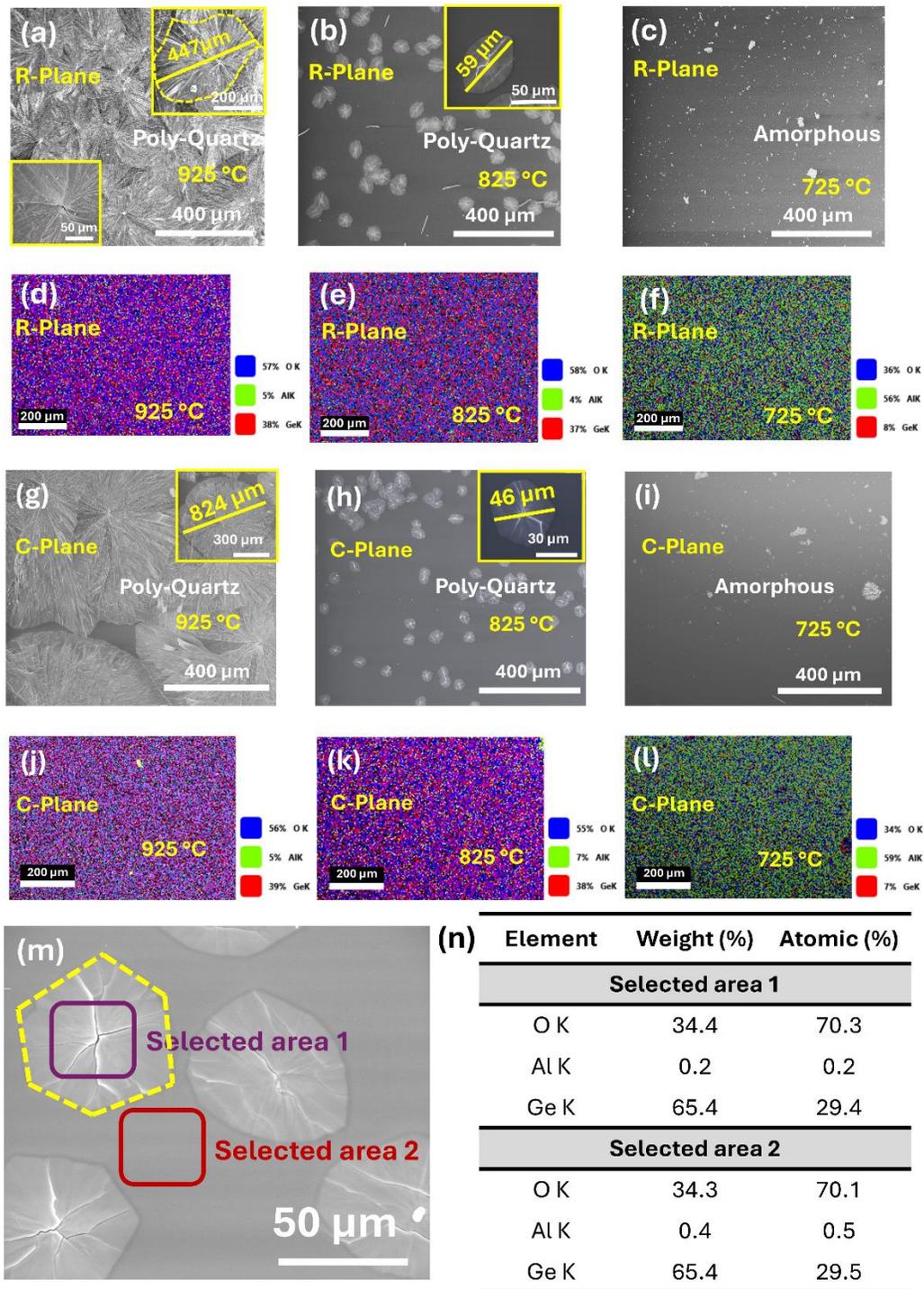

**Figure 2.** SEM images showing the impact of growth temperature on *R*-plane sapphire at (a) 925 °C, (b) 825 °C, and (c) 725 °C, with corresponding EDX analyses in (d-f). Growth temperature effects on *C*-plane sapphire are shown in (g) 925 °C, (h) 825 °C, and (i) 725 °C, with corresponding EDX analyses in (j-l). Comparison of chemical composition analysis by (m) SEM image and (n) EDX data for spherulites vs. non-spherulites (smooth) regions.

### 3.2 Effects of growth pressure

The effects of varying chamber pressure on the crystal structure of GeO$_2$ thin films are delineated in Figures 3 and 4. Figure 3(a) and (b) showcase the XRD patterns of GeO$_2$ thin films by varying the chamber pressure from 100 Torr to 300 Torr, grown on *R*-plane and *C*-plane sapphire substrates, respectively, with the growth parameters detailed in Table I for sample IDs E7, E5, and E6.

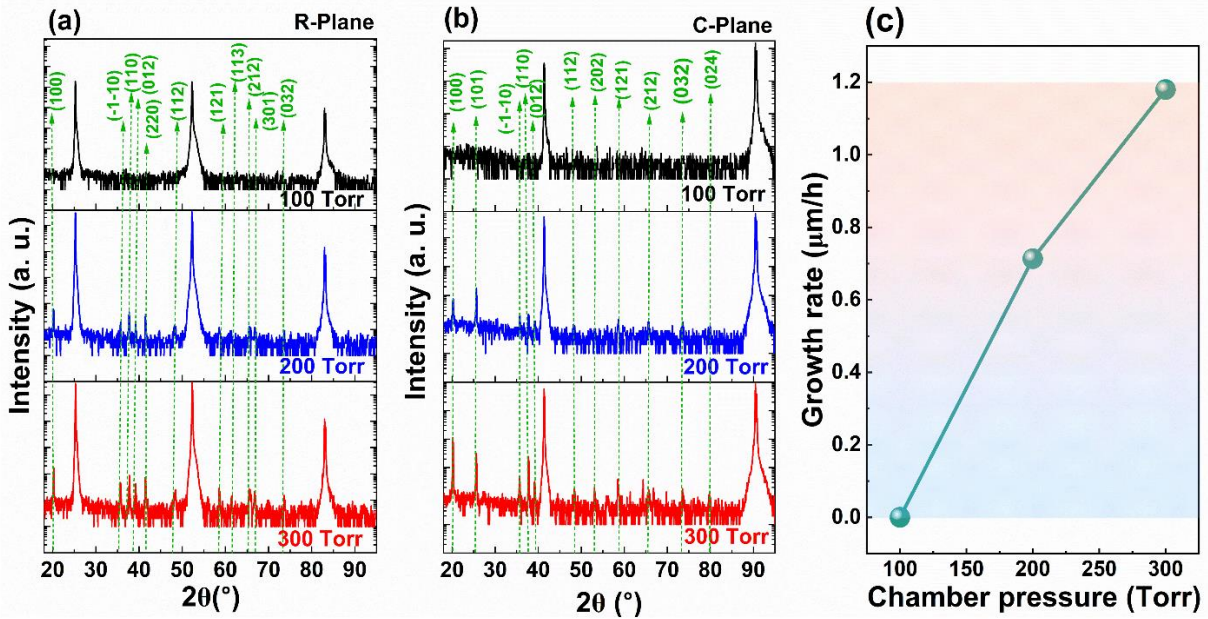

**Figure 3.** Impacts of different growth pressures, ranging from 100 to 300 Torr by XRD on (a) *R*-plane sapphire, (b) on *C*-plane sapphire substrate.(c) Growth rate as a function of chamber pressure.

The analysis reveals that there is no film at 100 Torr while the polycrystalline quartz phase was obtained at the higher growth pressures of 200 Torr and 300 Torr. In our study of the polycrystalline quartz phase, we observed distinct X-ray diffraction (XRD) peaks on films grown on *R*-plane sapphire substrates at the (100), (-1-10), (110), (012), (220), (112), (121), (113), (212), (301), and (032) planes. Conversely, the peaks at (101), (112), (202), (301), and (024) planes were exclusive to films on *C*-plane sapphire substrates. These additional peaks seen on the *C*-plane do not appear on the *R*-plane because they overlap with the distinguished peaks of the *R*-plane sapphire. Notably, the (220) plane is unique to the *R*-plane and this peak is absent on the *C*-plane due to the similar overlapping peak issue. Remarkably, neither substrate shows the characteristic peaks of the quartz GeO$_2$ phase at a chamber pressure of 100 Torr since the gas phase reactions and desorption are enhanced at low chamber pressures consuming the TMGe precursors. Figure 3(c) shows that the growth rate of the film increases with the increase of chamber pressure: 0 μm/h at 100 Torr, 0.71 μm/h at 200 Torr, and 1.18 μm/h at 300 Torr.

To understand the growth process, the possible equilibrium reactions [20] are shown in equations (1), and (2):

$$GeO_2(s) = GeO(g) + \frac{1}{2}O_2(g) \tag{1}$$

$$GeO_2(s) = GeO_2(g) \tag{2}$$

At lower pressures, decomposition becomes more pronounced, with the vapor phase primarily manifesting within this pressure regime, as detailed in reported phase diagrams [25]. Meanwhile, Wang *et al.* outline a mechanism by which germanium oxide (GeO) can desorb from amorphous $GeO_2$ at the $GeO_2$/Ge interface, involving the generation and diffusion of oxygen vacancies ($V_O$) to the surface, leading to the following reaction [26], [27], [28]:

$$GeO_2(s) + V_O \rightarrow GeO(g) \tag{3}$$

We anticipate the existence of $V_O$ at the $GeO_2/Al_2O_3$ interface, potentially augmenting the decomposition of $GeO_2$ films under low-pressure conditions, despite the films not being grown on a Ge substrate. The SEM images of the films are shown in Figures 4(a-c) for *R*-plane sapphire and 4(g-i) for *C*-plane sapphire. It is evident that pressure plays a crucial role in driving the transition from a film-less state to bulk film formation and further to fibrous films as pressure increases. The chemical composition analysis through SEM-EDX is shown below the corresponding SEM images, as detailed in Figures 4(d-f) for *R*-plane and 4(j-l) for *C*-plane sapphires. We found that at pressures lower than 100 Torr, no $GeO_2$ film was obtained for any chamber temperature, $O_2$ flow rate, and TMGe flow rate, which was confirmed by the EDX.

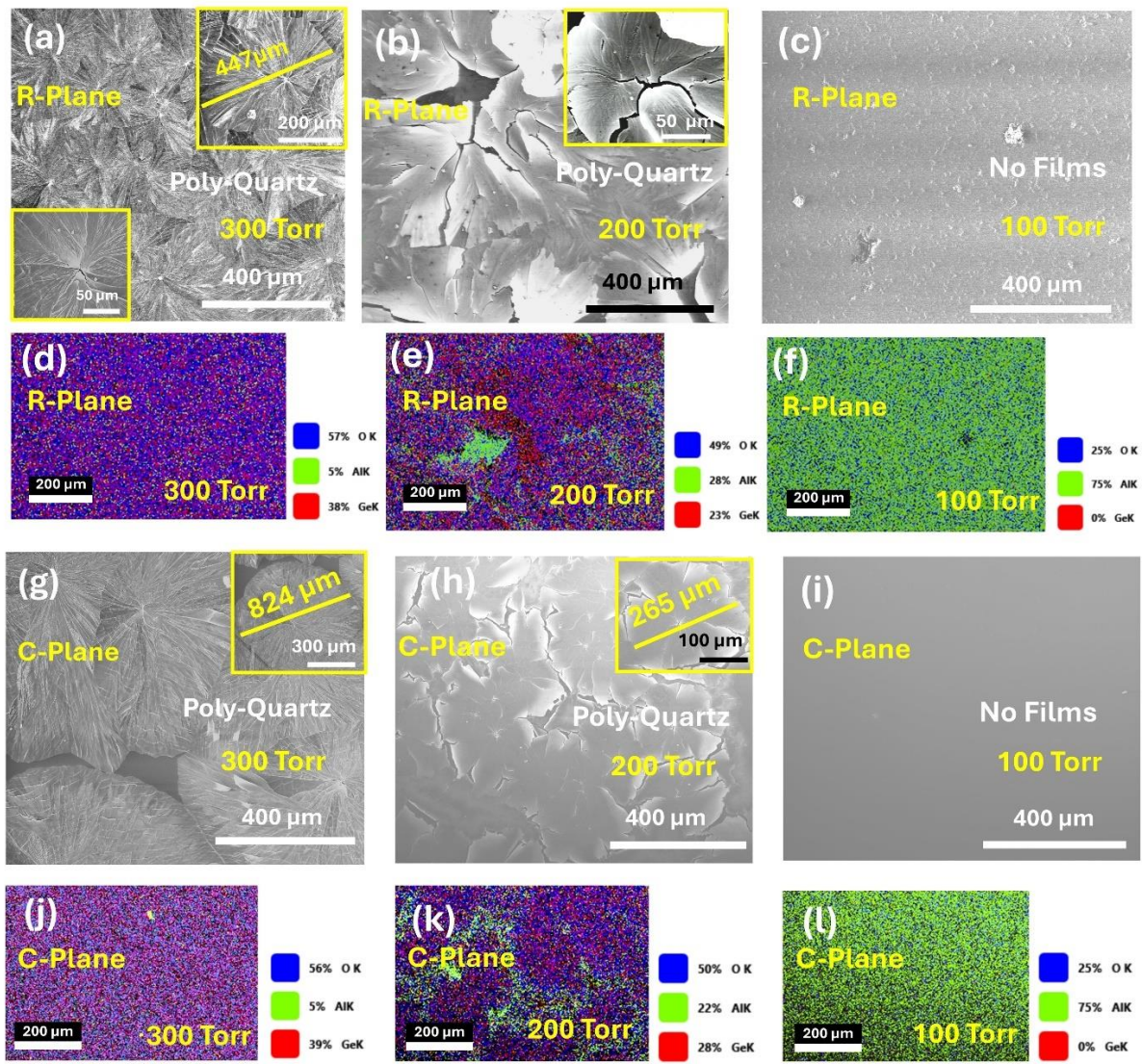

**Figure 4.** SEM images showing the impact of growth pressure on *R*-plane sapphire at (a) 300 Torr, (b) 200 Torr, and (c) 100 Torr, with corresponding EDX analyses in (d-f). Growth pressure effects on *C*-plane sapphire are shown in (g) 300 Torr, (h) 200 Torr, and (i) 100 Torr, with EDX analyses in (j-l).

### 3.3. Effects of Oxygen flow rate

The influence of oxygen flow rate on the crystal structure of $GeO_2$ thin films is examined in Figures 5 and 6. Specifically, Figure 5(a) and (b) present the X-ray diffraction (XRD) patterns for $GeO_2$ thin films, illustrating variations in oxygen flow rates from 450 SCCM to 1800 SCCM, while maintaining a constant TMGe precursor flow rate of 3.5 SCCM. These films were developed on both *R*-plane and *C*-plane sapphire substrates, adhering to the growth conditions outlined in Table I for sample IDs E7, E9, and E10, with both the growth temperature and pressure held steady at 925 °C and 300 Torr, respectively. Significantly, the analysis reveals that on *C*-plane sapphire, the (100) plane of $GeO_2$ exhibited a higher prominence compared to the (101) plane at oxygen flow rates of 450 SCCM and 1800 SCCM. However, at a flow rate of 900 SCCM, the intensities of the peaks for these two planes were equal. The difference in peak intensity is not due to film thickness or growth rate since the growth rate—1.06 µm/h at 450 SCCM, 1.12 µm/h at 900 SCCM, and 1.18 µm/h at 1800 SCCM—are very similar as depicted in Figure 5(c). The findings with differing oxygen flow rates can be attributed to several intertwined factors. The optimal oxygen flow rate of 900 SCCM, which equalizes the prominence of the (100) and (101) planes, might be providing a balanced environment that supports uniform surface kinetics and thermodynamics for both crystal orientations, possibly due to favorable gas phase dynamics and adsorption/desorption rates. In contrast, the flow rates of 450 SCCM and 1800 SCCM may not provide the same balance, potentially due to altered surface energy conditions or mass transport limitations, which could preferentially enhance the growth of the (100) plane. Despite the growth rates remaining relatively consistent across the different $O_2$ flow rates, indicating stable deposition, these subtle variations in flow are still significant enough to impact the resulting crystal structure.

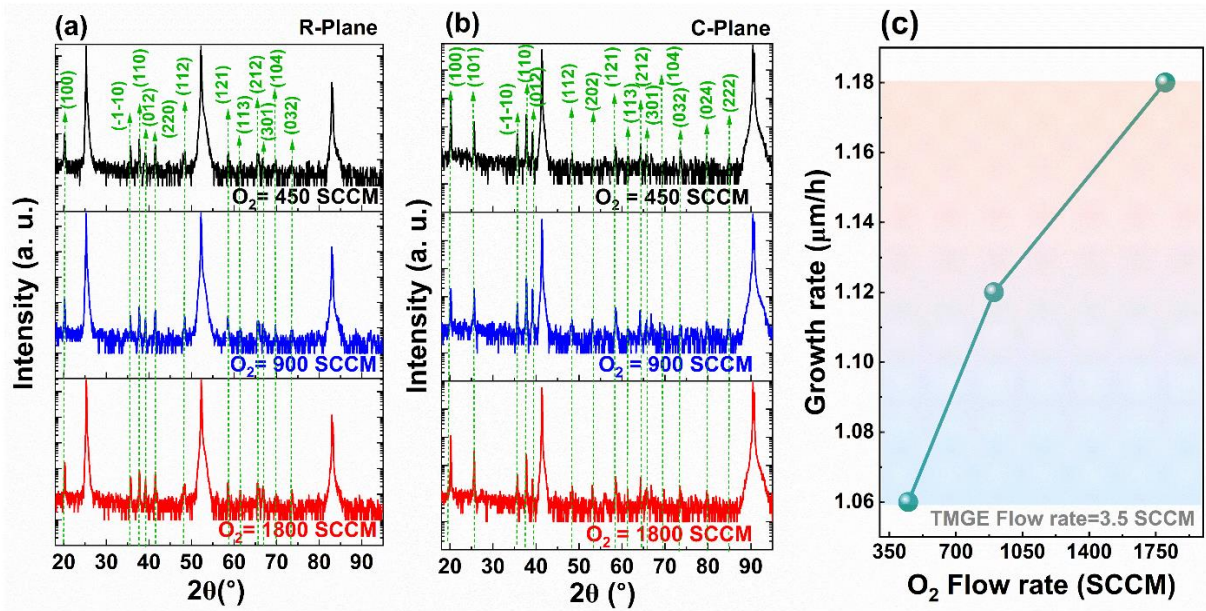

**Figure 5.** Influence of oxygen flow rate variation from 450 to 1800 SCCM on XRD patterns, with a constant precursor flow rate of 3.5 SCCM, for (a) *R*-plane and (b) *C*-plane sapphire substrates. (c) Growth rate as a function of $O_2$ flow rate.

The SEM images illustrate similar spherulite patterns, as seen in Figures 6(a)-(c) for *R*-plane sapphire and 6(g)-(i) for *C*-plane sapphire substrates. Additional chemical composition insights obtained through SEM-EDX are shown below the corresponding SEM images, detailed in Figures 6(d)-(f) for *R*-plane and 6(j)-(l) for *C*-plane sapphires. It is apparent that changes in the oxygen flow rate do not affect the polycrystalline patterns. Nonetheless, a higher oxygen flow rate assists in suppressing decomposition, as evidenced by equations (1)-(3), thereby resulting in an increased growth rate.

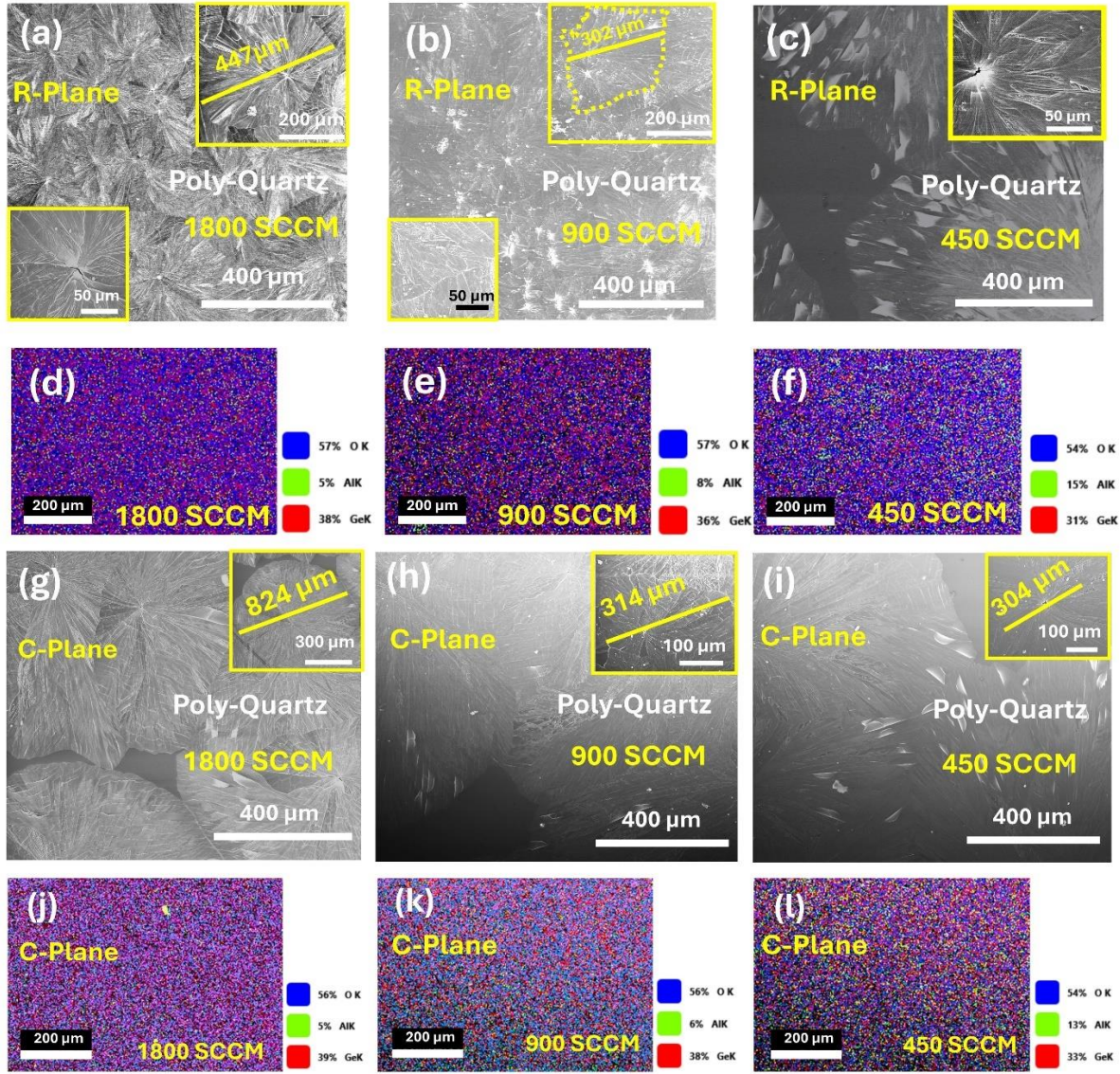

**Figure 6.** SEM images showing the impact of oxygen flow rate, keeping the precursor flow rate at 3.5 SCCM on *R*-plane sapphire at (a) 1800 SCCM, (b) 900 SCCM, and (c) 450 SCCM, with corresponding EDX analyses in (d-f). $O_2$ flow rate effects on *C*-plane sapphire are shown in (g) 1800 SCCM, (h) 900 SCCM, and (i) 450 SCCM, with EDX analyses in (j-l).

### 3.4. Effects of shroud gas flow rate

The effect of varying shroud gas (argon, Ar) concentrations on the surface morphology of $GeO_2$ thin films is thoroughly investigated in Figures 7 and 8. Figures 7(a) and (b) display XRD patterns of $GeO_2$ thin films and 7(c) shows the growth rate as a function of shroud gas flow rate, capturing the impact of shroud gas flow rate ranging from 400 SCCM to 2800 SCCM, while keeping the TMGe precursor flow rate at 3.5 SCCM and oxygen level at 900 SCCM constant. These films, grown on both *R*-plane and *C*-plane sapphire substrates, follow the growth parameters specified in Table I for sample IDs E9, E12, and E15, with the growth temperature and pressure maintained at 925 °C and 300 Torr, respectively. There is no noticeable change in peak intensity with the changing shroud gas flow rate from 2800 SCCM to 400 SCCM for both *C*-plane and *R*-plane sapphire substrates. However, For the *C*-plane sapphire, the (100) crystal plane was more prominent than the (101) plane at shroud gas flow rates of 400 SCCM and 1400 SCCM. Yet, at a higher flow rate of 2800 SCCM, both the (100) and (101) planes showed equal peak intensities. The variation in crystal quality of $GeO_2$ films grown by MOCVD on sapphire substrates, with changes in shroud gas flow rate, suggests a nuanced relationship between gas dynamics and film structure. Specifically, at lower flow rates (400 and 1400 SCCM), the conditions seem to favor the formation of the (100) plane on the *C*-plane sapphire, possibly due to lower interaction rates of gas molecules with the surface, which may promote faster nucleation or stabilization of this plane. Conversely, at the highest tested flow rate (2800 SCCM), the increased interaction and possibly altered stoichiometry enable the (101) plane to form with equal intensity, indicating that such conditions create a thermodynamic or kinetic environment that does not preferentially support one crystal plane over another. The growth rate, showing minimal variation from 1.02 μm/h at 400 SCCM to 1.12 μm/h at 2800 SCCM, points to the dominance of other growth-controlling mechanisms besides shroud gas flow rate, such as surface reactions or inherent material properties.

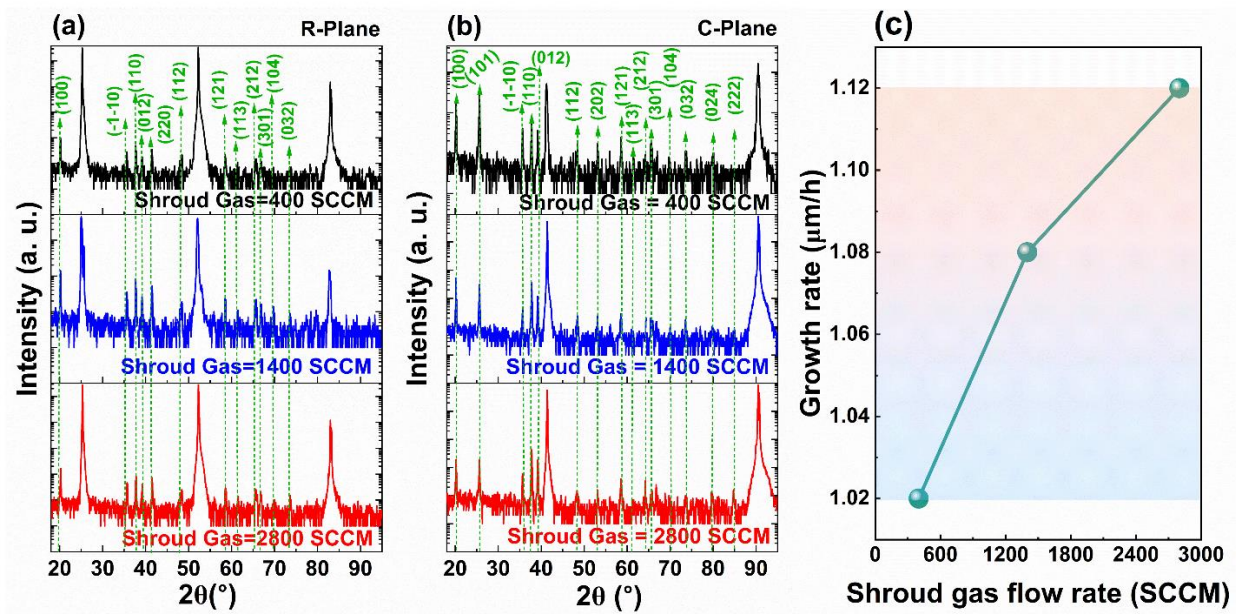

**Figure 7.** Influence of shroud gas flow rate from 400 to 2800 SCCM on XRD patterns, for (a) *R*-plane and (b) *C*-plane sapphire substrates. (c) Growth rate as a function of shroud gas flow rate.

The SEM images clearly showcase the similar spherulites on *R*-plane and *C*-plane sapphire substrates, as depicted in Figures 8(a-c) and 8(g-i), respectively. Further insights into the chemical composition are revealed by SEM-EDX analyses in Figures 8(d-f) for the *R*-plane and 8(j-l) for the *C*-plane sapphires. It is evident that a low flow rate (400 SCCM) of the shroud gas fosters the formation of spherulites, while higher flow rates (1400 SCCM and 2800 SCCM) facilitate the emergence of large, spherulitic polycrystalline patterns.

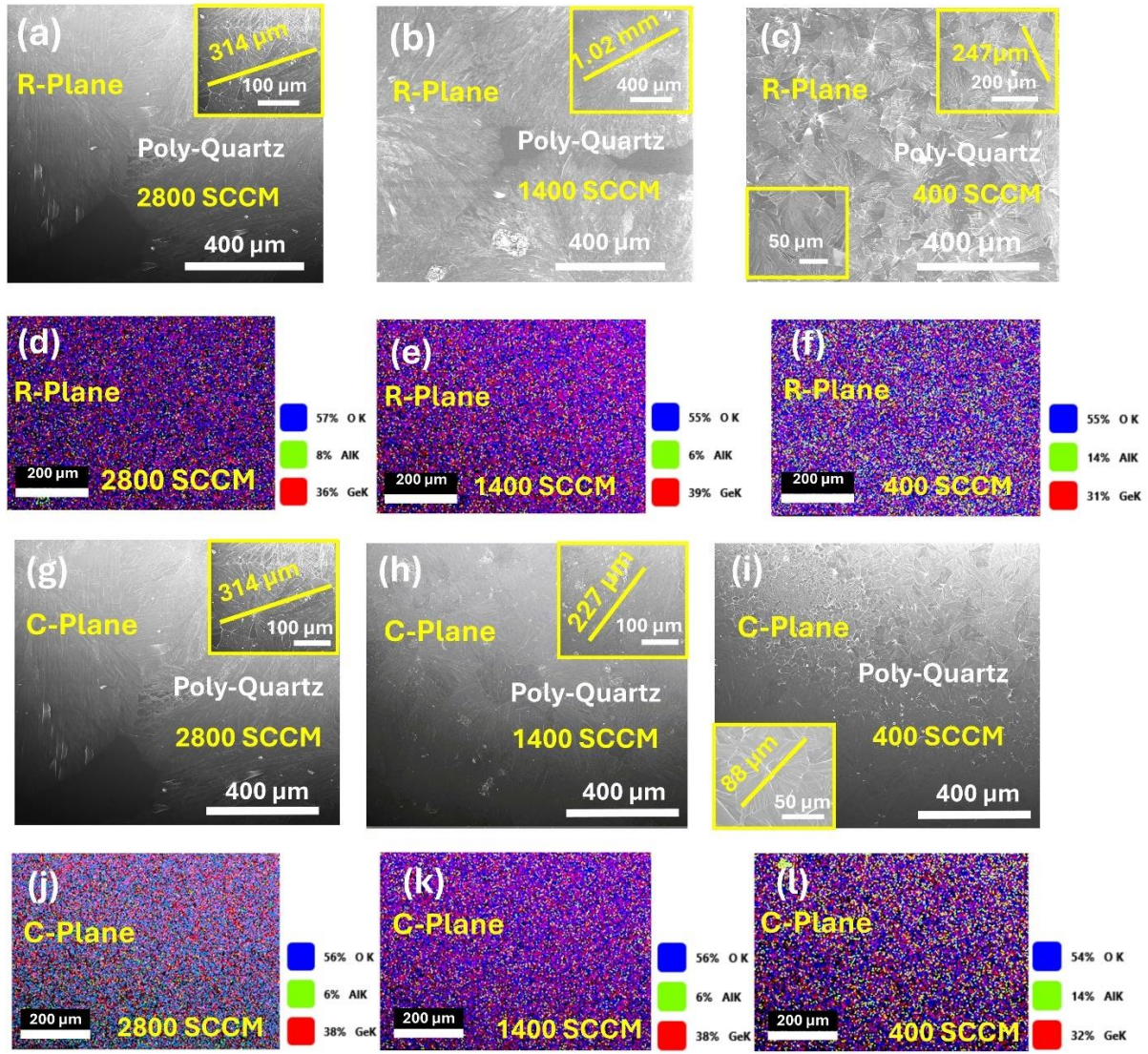

**Figure 8.** SEM images showing the impact of shroud gas flow rate, keeping the precursor flow rate at 3.5 SCCM on *R*-plane sapphire at (a) 2800 SCCM, (b) 1400 SCCM, and (c) 400 SCCM, with corresponding EDX analyses in (d-f). Shroud gas flow rate effects on *C*-plane sapphire are shown in (g) 2800 SCCM, (h) 1400 SCCM, and (i) 400 SCCM, with EDX analyses in (j-l).

### 3.5. Effects of TMGe precursor flow rate

The TMGe precursor flow rate significantly influences the thickness of the film, with a higher rate correlating to more pronounced intensities of the polycrystalline quartz peaks, as illustrated in Figures 9(a) and (b). The specific growth conditions associated with these observations are meticulously cataloged in Table I, under sample IDs E1, and E7.

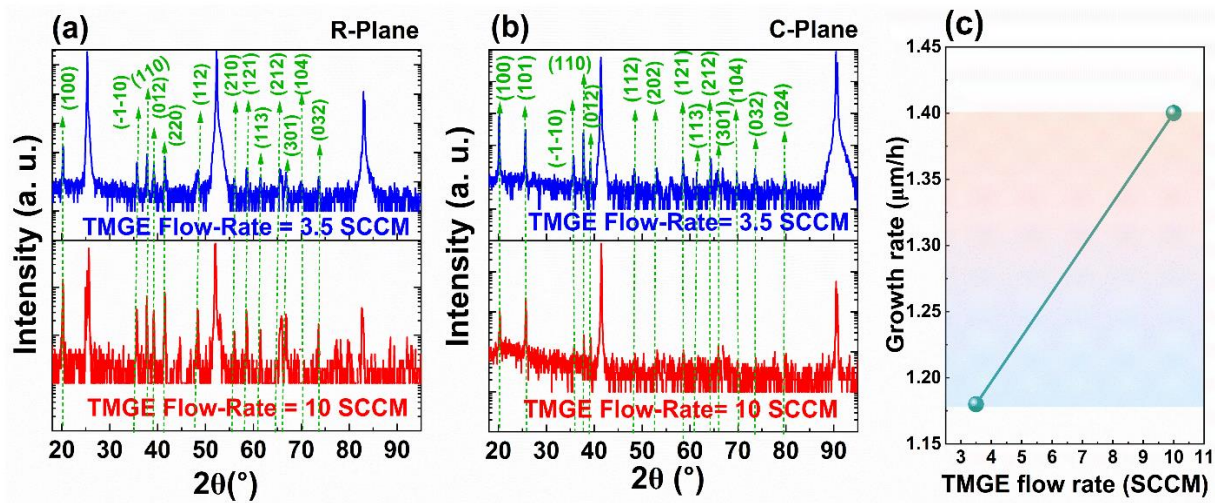

**Figure 9.** Influence of TMGe precursor flow rate variation from 1.5 to 3.5 SCCM on XRD patterns, with a constant oxygen flow rate of 900 SCCM, for (a) *R*-plane and (b) *C*-plane sapphire substrates. (c) Growth rate as a function of TMGe flow rate.

Figure 9(c) indicates the growth rates for TMGe precursor flow rates at 3.5 SCCM and 10 SCCM, measurably 1.18 µm/h and 1.4 µm/h, respectively. The distinct crystal plane prominence on the *C*-plane sapphire— (101) over (100) at 10 SCCM, and the opposite at 3.5 SCCM—could be due to how varying TMGe precursor flow rates affect the surface chemistry and reaction kinetics, leading to changes in the preferential growth of specific crystal planes. The SEM images highlight the same surface morphology of large spherulite quartz films on both *R*-plane and *C*-plane sapphire substrates, as shown in Figures 10(a-b) and 10(e-f). Chemical analyses via SEM-EDX are shown below the corresponding SEM images in Figures 10(c)-(d) for the *R*-plane and 10(g)-(h) for *C*-plane sapphires. It is apparent that elevating the flow rate of the TMGe precursor prompts the transformation of $GeO_2$ films from spherulitic polycrystalline patterns to bulk film formations.

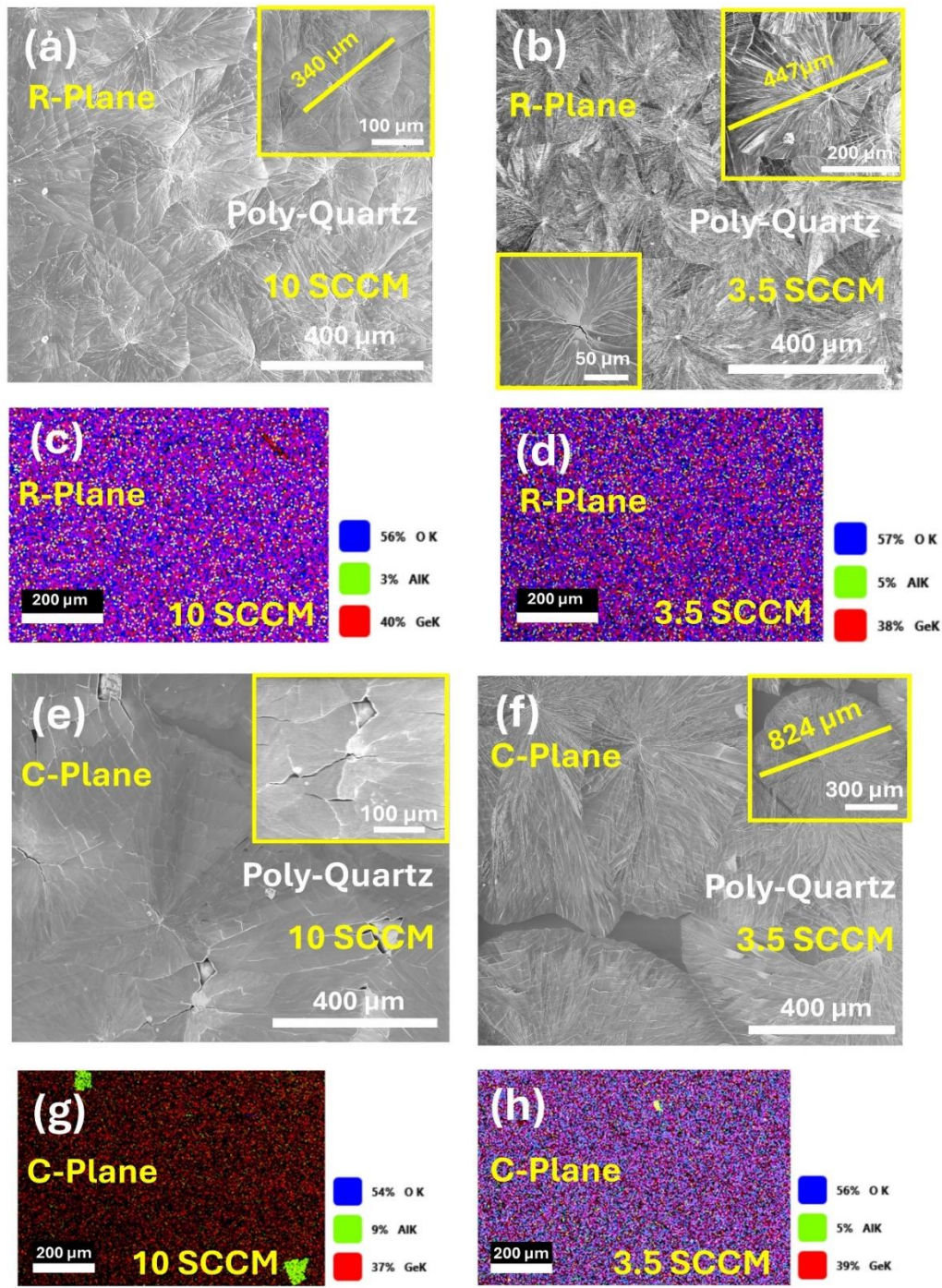

**Figure 10.** SEM images showing the impact of TMGe precursor flow rate, keeping the precursor flow rate at 3.5 SCCM on *R*-plane sapphire at (a) 10 SCCM, and (b) 3.5 SCCM, with corresponding EDX analyses in (c-d). TMGe precursor flow rate effects on *C*-plane sapphire are shown in (e) 10 SCCM, and (f) 3.5 SCCM with EDX analyses in (g-h).

### 3.6. Effects of susceptor rotation speed

Figures 11(a) and (b) present the XRD patterns of films grown on both *R*-plane and *C*-plane sapphire substrates at rotation speeds of 2 RPM and 300 RPM. The growth conditions for these observations are meticulously cataloged in Table I, under sample IDs E1 and E16 for the respective rotation speeds of 300 RPM and 2 RPM. The data clearly illustrate that at a high rotation speed of 300 RPM, the film exhibits polycrystalline quartz characteristics. Conversely, at a much lower speed of 2 RPM, the film remains amorphous [29], [30].

$$\delta \sim \sqrt{\frac{\mu H}{\rho v d^2}} \tag{4}$$

where $\mu$ is the dynamic viscosity, H is the reactor height, $\rho$ is the gas density, $v$ is the flow velocity, and $d$ is the chamber diameter. Higher rotation speed could thinner the boundary layer and increase the growth rate.

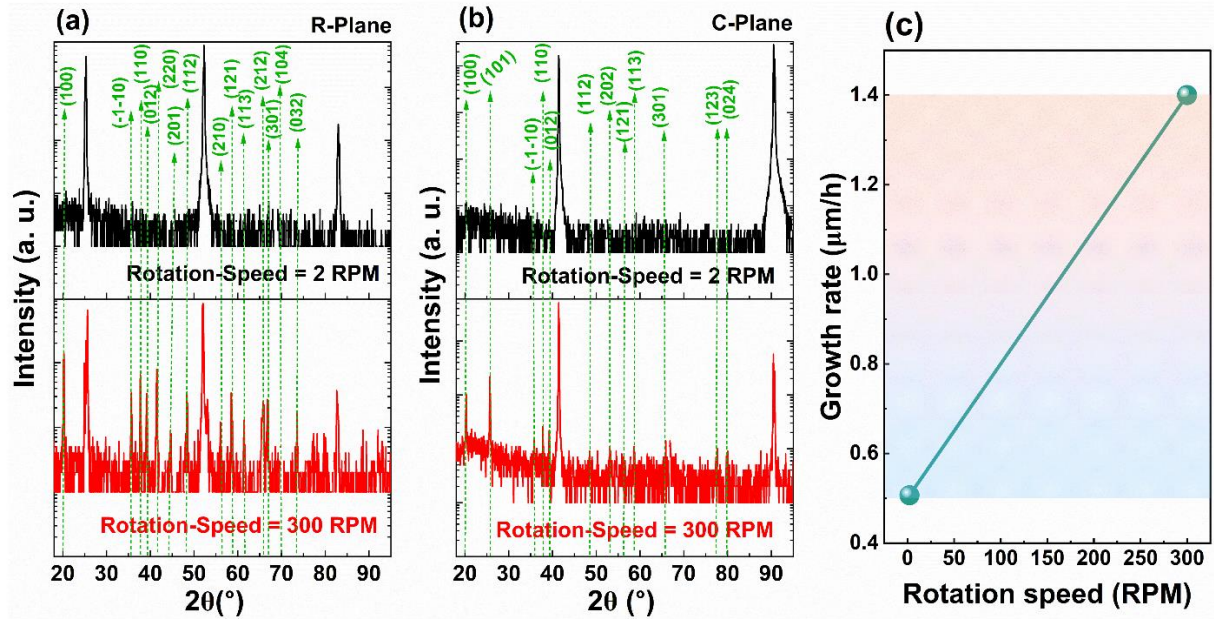

**Figure 11.** Impacts of different susceptor rotation speeds from 2 RPM to 300 RPM by XRD on (a) *R*-plane sapphire, (b) on *C*-plane sapphire substrate. (c) Growth rate as a function of susceptor rotation speed.

At a high rotation speed, such as 300 RPM, the thickness of this thermal boundary layer is significantly reduced, which promotes the formation of polycrystalline quartz films. Conversely, at a slow speed of 2 RPM, a thicker thermal boundary layer is maintained, leading to the film remaining amorphous with a much slower growth rate of 0.51µm/h. These SEM images of the samples are depicted in Figures 12(a) and (b) for the R-plane and 12(e) and (f) for the C-plane, while the EDX results are shown in Figs. 12(c) and (d), and 12(g) and (h), respectively.

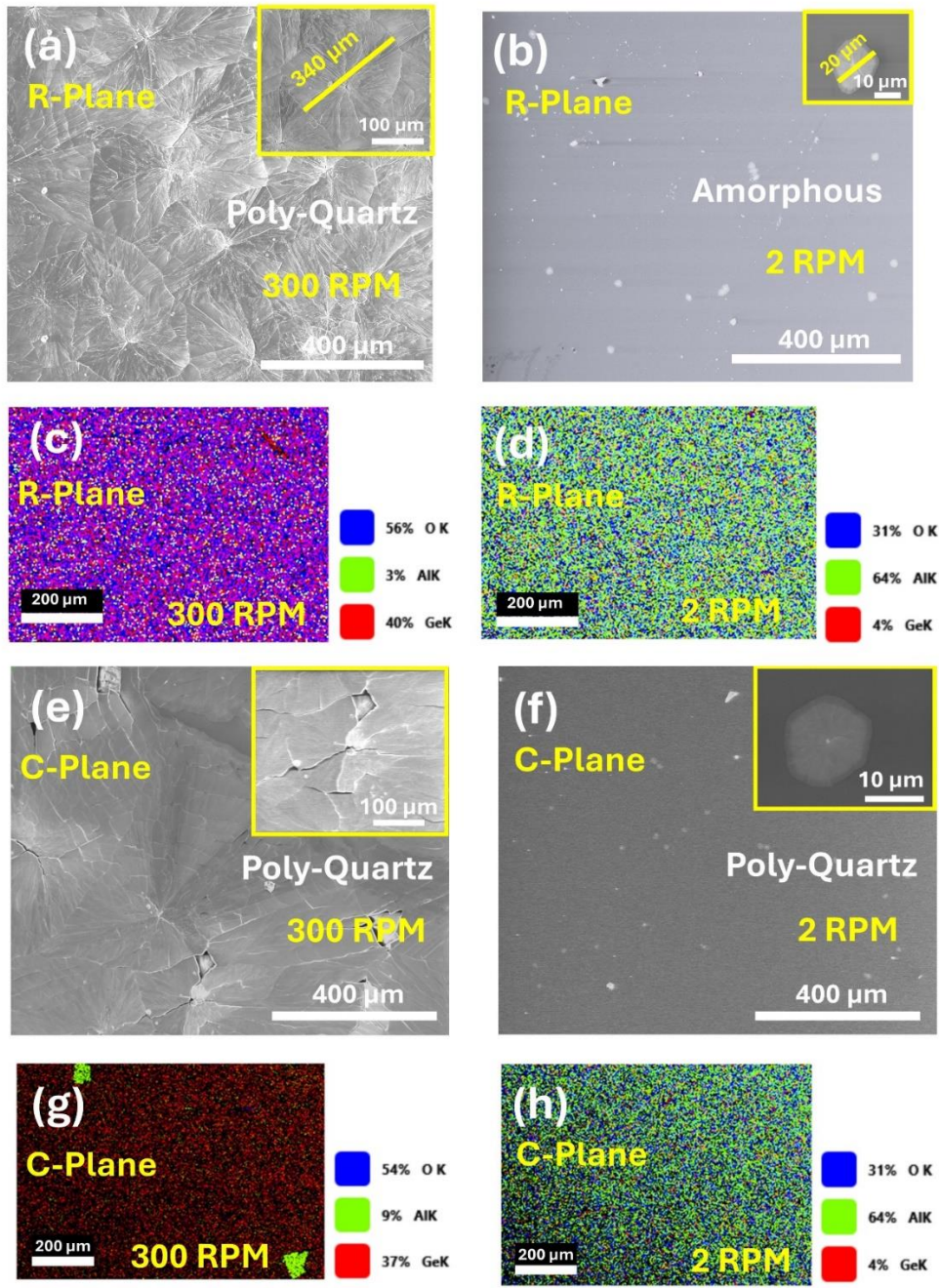

**Figure 12.** SEM images show the impact of susceptor rotation speed on *R*-plane sapphire at (a) 300 RPM, and (b) 2 RPM, with corresponding EDX analyses in (c-d). Influences of susceptor rotation speed on *C*-plane sapphire are shown in (e) 300 RPM, and (f) 2 RPM with EDX analyses in (g-h).

In addition, the root mean square (RMS) surface roughness of the film, growth at 925 °C (sample ID: E7) has been quantitatively assessed using Atomic Force Microscopy (AFM). Due to the variable sizes of the spherulites, achieving uniform measurements of surface roughness presented challenges in certain instances. Nevertheless, the AFM analysis provides an overview of the RMS roughness of the film in the non-spherulite region, which is approximately 10.27 nm. Additionally, the height profile depicted in Figure 13, illustrates the discrepancy between the smooth areas of the film and those adorned with spherulites, indicating a height difference of about 13 nm.

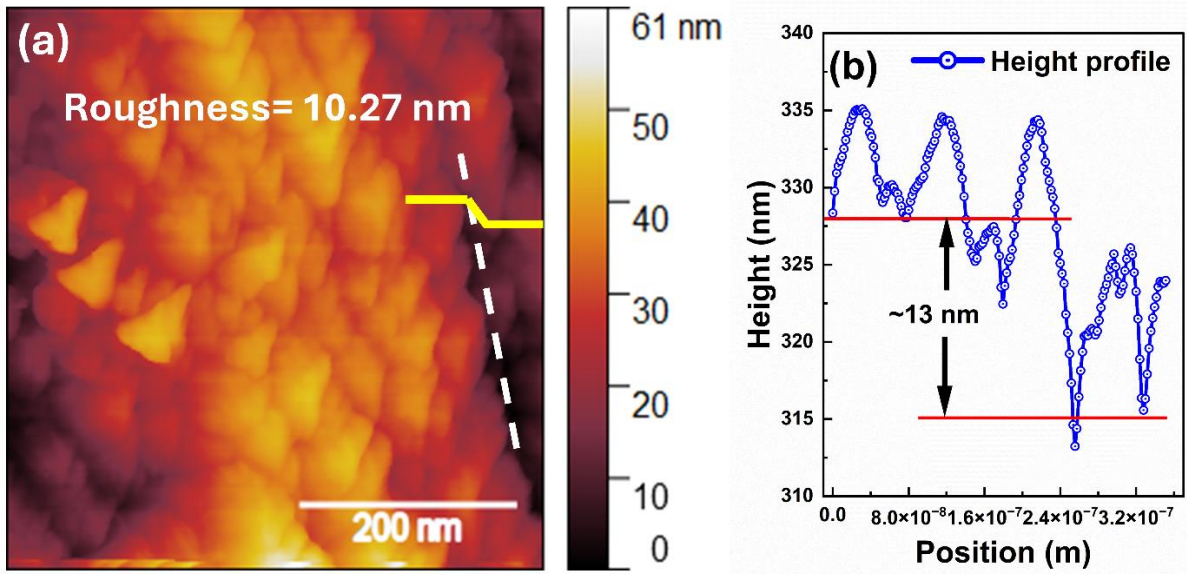

**Figure 13.** AFM surface morphology of thin film surface, showing (a) RMS roughness, (b) height profile.

A phase diagram is created shown in Fig. 14, summarizing the experimental findings for the growth of $GeO_2$ thin films via MOCVD using the TMGe precursor. It includes a legend with brown, ash, and red stars representing experimental observations: 'No Films,' 'Amorphous,' and 'Polycrystalline' structures, respectively. The ash region denotes the amorphous phase of the thin films, while the brown region at the bottom signifies conditions under which no films were formed, as evidenced by the brown stars. The presence of the polycrystalline phase is indicated by red stars within the specified growth region. The white or unmarked areas represent regions that need further exploration to determine their crystallinity. This diagram provides a visual hypothesis of the phase distribution based on the observed experimental results.

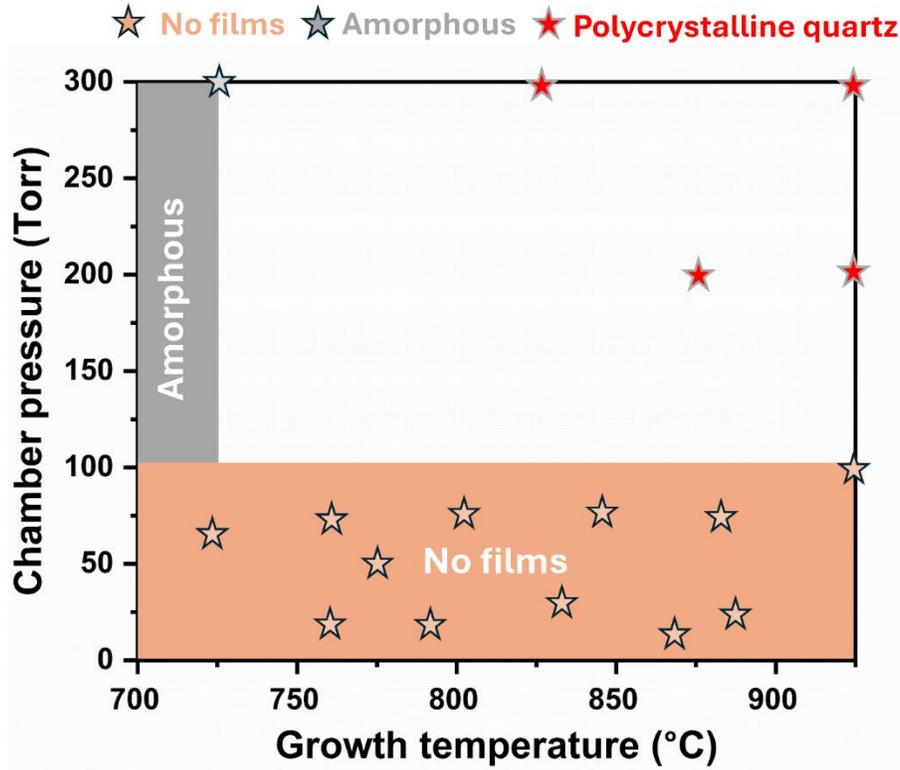

**Figure 14.** Phase diagram based on the experimental results.

## 4. Conclusion

In this study, we conducted an extensive investigation into the growth of GeO$_2$ films via MOCVD, meticulously examining how variations in key growth parameters affect film quality. By adjusting growth temperature, pressure, oxygen concentration, precursor flow rate, shroud gas composition, and rotation speed across both R- and *C*-plane substrates, we have identified the transition process of GeO$_2$ films from a filmless state to an amorphous phase and to a polycrystalline phase. A phase diagram has been constructed based on the experimental results, outlining the growth windows for GeO$_2$ films on sapphire substrates. This research serves as a pioneering guide for the MOCVD growth of GeO$_2$ films.

**Acknowledgments**

The authors thank the support from the University of Utah start-up fund.

**AUTHOR DECLARATIONS**
**Conflict of Interest**

The authors have no conflicts to disclose.

**DATA AVAILABILITY**

The data that support the findings of this study are available from the corresponding authors upon reasonable request.